\documentclass[%
reprint,
two-column,
superscriptaddress,
 amsmath,amssymb,
 aps, physrev,
]{revtex4-2}

\usepackage{graphicx}
\usepackage{dcolumn}
\usepackage{bm}

\usepackage{subfigure}

\usepackage{chemformula} 
\usepackage[T1]{fontenc} 
\usepackage{amssymb}
\usepackage{braket}
\usepackage{float}

\usepackage{ulem}

\usepackage{graphicx}
\usepackage{dcolumn}
\usepackage{bm}

\begin{document}

\preprint{}

\title{\textbf{Phonon scattering mechanisms in WTe$_{2}$ observed by ultrafast coherent phonon spectroscopy} 
}%
\author{Mizuki Akei}
\email{s2630089@u.tsukuba.ac.jp}
\affiliation{Department of Applied Physics, Graduate school of Pure and Applied Sciences, University of Tsukuba, 1-1-1 Tennodai, Tsukuba 305-8573, Japan}
\author{Yu Mizukoshi}
\affiliation{Department of Applied Physics, Graduate school of Pure and Applied Sciences, University of Tsukuba, 1-1-1 Tennodai, Tsukuba 305-8573, Japan}
\author{Muneaki Hase}
\email{mhase@bk.tsukuba.ac.jp}
\affiliation{Department of Applied Physics, Graduate school of Pure and Applied Sciences, University of Tsukuba, 1-1-1 Tennodai, Tsukuba 305-8573, Japan}


\date{\today}

\begin{abstract}
Revealing the mechanisms of phonon scattering is crucial for understanding material properties such as transport characteristics and optical responses. It can be discussed by measuring the temperature dependence of the phonon energy and lifetime. To gain insight into these mechanisms in Weyl semimetal $T_d$-WTe$_2$, we investigated coherent phonons using time-resolved pump-probe spectroscopy in a wide temperature range from 4.6 to 300 K. The temperature dependence of the frequency and decay rate of the two high-frequency optical modes was described by the conventional anharmonic phonon-phonon scattering model. In contrast, the low-frequency mode at 2.4 THz exhibited anomalous behavior, which can be interpreted as the contribution of phonon-electron scattering.
\end{abstract}

\maketitle

Weyl semimetals are types of topological materials that have attracted significant interest due to their intriguing physical properties and broad potential applications such as terahertz frequency-driven optoelectronic memory devices \cite{xiao2020berry}. Especially, WTe$_2$ has recently provided an ideal platform for exploring novel physical properties and controlling their characteristics. $T_d$-WTe$_2$ is classified as a type-II Weyl semimetal, characterized by a tilted Weyl cone and forming electron and hole pockets \cite{soluyanov2015type}. Based on this distinctive electronic structure, WTe$_2$ exhibits unique physical properties, such as large non-saturating magnetoresistance \cite{ali2014large, thoutam2015temperature}, temperature-induced Lifshitz transition \cite{wu2015temperature,zhang2017lifshitz}, pressure-induced superconductivity \cite{pan2015pressure}, and photoinduced phase transition \cite{sie2019ultrafast,Akei2025switching}. In particular, a large non-saturating magnetoresistance at low temperature is attributed to an almost perfect compensation between electrons and holes \cite{ali2014large}. In addition, anisotropic magnetoresistance was observed below $\sim$ 75 K, indicating changes in the electronic structure with temperature \cite{thoutam2015temperature}. 

Further evidence of changes in the electronic structure at low temperatures was provided by studies of ultrafast carrier dynamics. It was found that the phonon-assisted electron-hole recombination process between the electron and hole pockets gradually slow down with the cooling temperature as the phonon density decreases, but becomes faster again below $\sim$ 50 K \cite{dai2015ultrafast}. Moreover, angle-resolved photoemission spectroscopy measurements revealed that as the temperature increases, the size of the hole pocket decreases due to a shift in the chemical potential, and eventually the hole pocket disappears around 160 K \cite{wu2015temperature}. This is called the Lifshitz transition, a change in the topology of the Fermi surface driven by external forces such as temperature, pressure, and carrier doping, and does not necessarily require a phase transition accompanied by a breaking of symmetry \cite{lifshitz1960anomalies}. On the other hand, other studies indicate that the two hole pockets subsequently disappear at $\sim$ 90 K and $\sim$ 220 K, respectively \cite{zhang2017lifshitz}.  
Thus, the unique temperature dependence of the electronic states at low temperatures has been controversial. Furthermore, it remains unclear how changes in the electronic state affect phonon scattering, and the study of phonon dynamics in WTe$_2$ is limited \cite{kong2015raman}. Therefore, investigating phonon-phonon and electron-phonon interactions could provide important insights into the resolution of these issues. To this end, temperature-controlled coherent phonon spectroscopy provides a powerful tool for exploring the electron-phonon interaction.

In this paper, we explore the phonon scattering processes in WTe$_2$ by optical pump-probe coherent phonon spectroscopy with varying temperature from 4.6 K to 300 K. We observed seven $A_1$ optical phonon modes and evaluated the temperature dependence of frequencies and decay rates, focusing on representative intralayer optical modes with large amplitudes.  
The temperature dependence of the two high-frequency modes exhibits conventional behavior and can be explained by the anharmonic phonon-phonon scattering model. In contrast, the low-frequency mode at 2.4 THz exhibits anomalous temperature dependence that cannot be described by anharmonic phonon-phonon scattering. This behavior is attributed to the contribution of interband phonon-electron scattering, in which phonons decay into electron-hole pairs.

We carried out measurements of the transient reflectivity change ($\Delta R/R$) based on the pump-probe method. As a light source, we used a Ti: sapphire oscillator with a central wavelength of 830 nm, a pulse duration of $\sim 30$ fs, and a repetition rate of 80 MHz. The pump fluence was 10 $\mu$J/cm$^{2}$ (varying from 3 to 100 $\mu$J/cm$^{2}$ for fluence dependent measurements), weak enough to prevent photoinduced phase transitions and sample damage \cite{Akei2025switching}. A probe beam with fluence one order of magnitude lower than the pump beam was used. The time delay between the pump and probe pulses was scanned at 9.5 Hz by a shaker. The sample used in this study was a bulk single crystal of (001) surface $T_d$-WTe$_2$ (from HQ Graphene) with a thickness of $\sim$ 100 $\mu$m. It was mounted in a closed-cycle He gas cryostat maintained under high vacuum ($\sim1.2\times10^{-5}$ Pa), allowing precise control of temperatures from 4.6 to 300 K.

\begin{figure}
\makeatletter
\renewcommand{\thefigure}{%
\thesection \arabic{figure}}
\@addtoreset{figure}{section}
\makeatother
	\includegraphics[width=8.6cm]{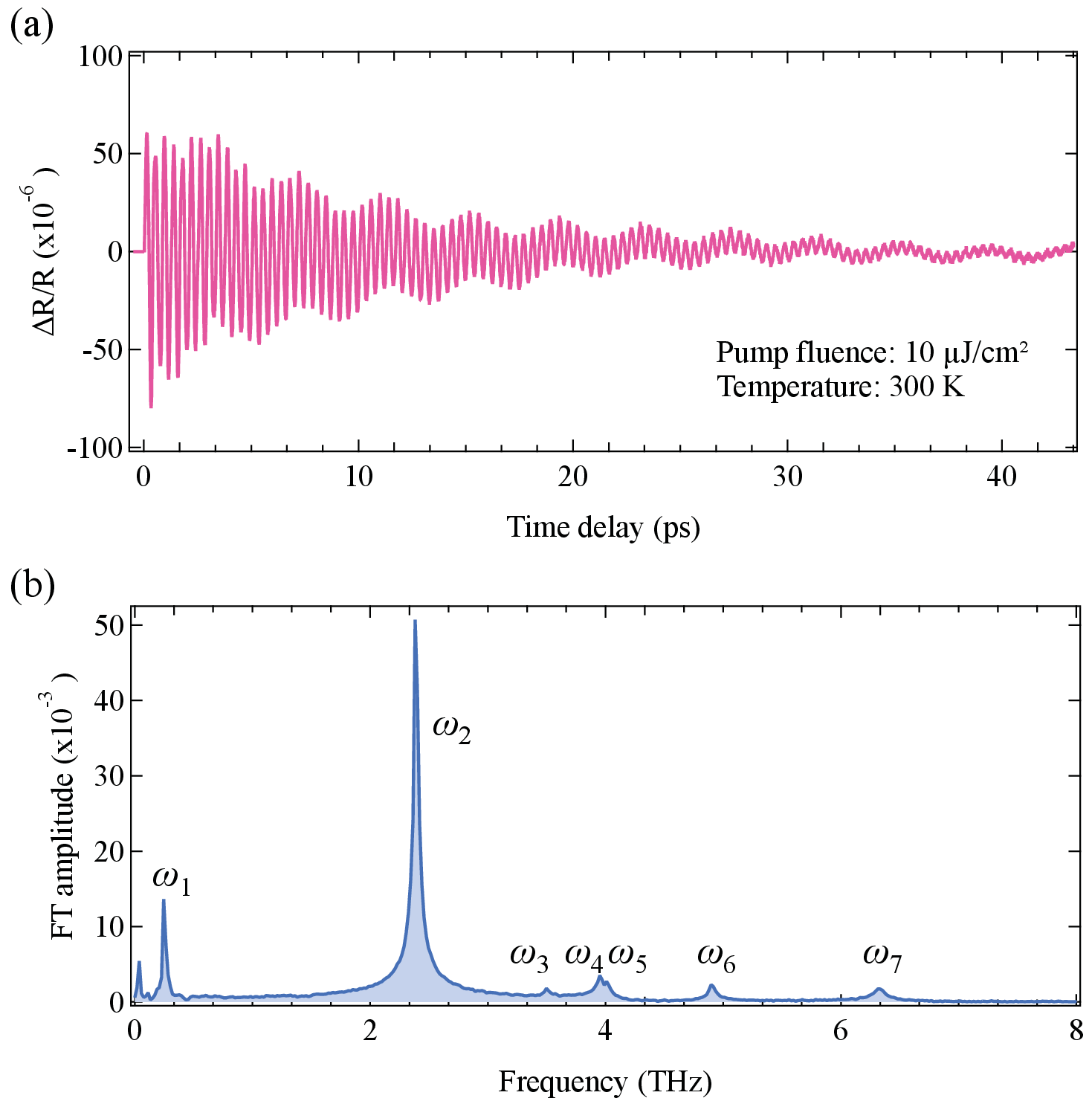}
	\caption{(a) The time-domain coherent phonon signal obtained by reflective pump-probe spectroscopy at 300 K with an excitation fluence of 10 $\mu$J/cm$^{2}$. (b) Fourier transform spectrum of (a). Seven optical $A_1$ phonon modes were observed. }
	\label{CPsignal}
\end{figure}

\begin{figure}
\makeatletter
\renewcommand{\thefigure}{%
\thesection \arabic{figure}}
\@addtoreset{figure}{section}
\makeatother
	\includegraphics[width=8.6cm]{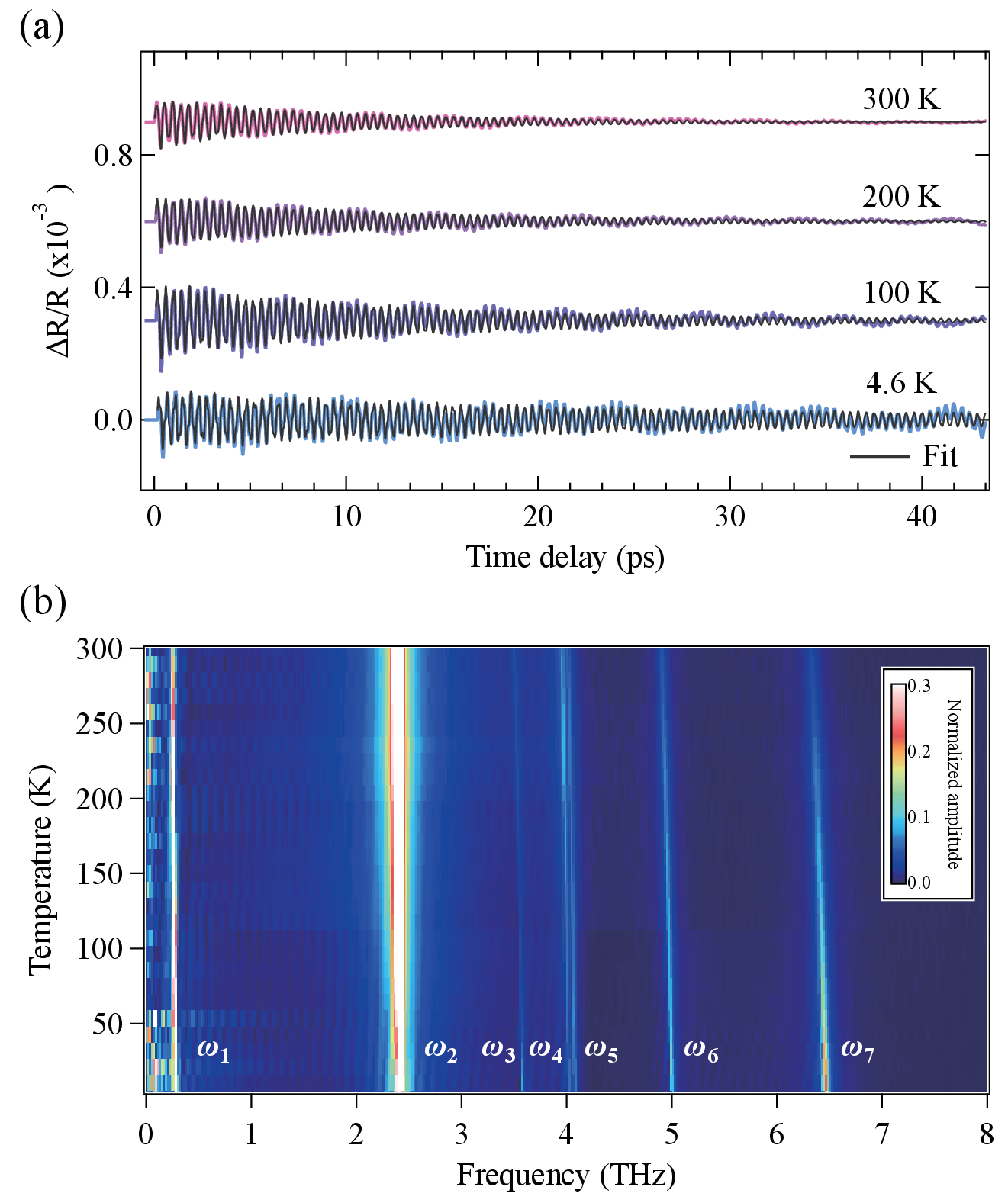}
	\caption{(a) The coherent phonon signals of $T_d$-WTe$_2$ measured with the excitation fluences of 10 $\mu$J/cm$^{2}$ at 4.6 K, 100 K, 200 K, and 300 K. The solid lines represent fit results using Eq. (\ref{damped}). (b) A two-dimensional color map of the Fourier transform spectrum of coherent phonon signals for each temperature. The vertical and horizontal axes represent temperature and frequency, respectively. }
	\label{temperature_dependence}
\end{figure}

Figure \ref{CPsignal}(a) shows time-domain coherent phonon signal at 300 K with an excitation fluence of 10 $\mu$J/cm$^{2}$, where the non-oscillatory component was removed by subtracting the exponential fitting of the carrier response \cite{dai2015ultrafast}. The corresponding Fourier transform (FT) spectrum is shown in Fig. \ref{CPsignal}(b), where peaks for seven phonon modes were observed. Each of the phonon frequencies was labeled as $\omega_1\sim\omega_7$ from lowest to highest frequency. At 300 K, these frequencies were $\omega_1=0.25\,\mathrm{THz},\,\omega_2=2.38\,\mathrm{THz},\,\omega_3=3.50\,\mathrm{THz},\,\omega_4=3.95\,\mathrm{THz},\,\omega_5=4.01\,\mathrm{THz},\,\omega_6=4.90\,\mathrm{THz},$ and $\omega_7=6.31\,\mathrm{THz}$. 
According to previous studies on Raman spectroscopy, all of these vibrational modes correspond to the $A_1$ optical phonon mode \cite{ma2016raman,jiang2016raman,song2016plane,kong2015raman}. In particular, $\omega_1$ is an interlayer shear mode, while $\omega_2 \sim \omega_7$ are intralayer optical phonon modes. These results are consistent with previous studies on coherent phonon spectroscopy \cite{soranzio2022strong,hein2020mode,drueke2021observation}. 
The FT peaks below 0.25 THz can originate from any instrumental noise since the peaks appear randomly at different temperatures (as visible in Fig. \ref{temperature_dependence}), and does not affect the analysis of the phonon modes.

To clarify the behavior of phonons as a function of temperature, we measured transient changes in reflectivity by varying the temperature from 4.6 to 300 K with a pump fluence of 10 $\mu$J/cm$^{2}$. 
In order to minimize the effects of the lattice and electron temperature rises and observe the net temperature effects, we discuss the phonon scattering mechanisms using temperature dependent measurements at the ultralow pump fluence of 10 $\mu$J/cm$^{2}$ (the details are in Appendix A). Figure \ref{temperature_dependence}(a) shows the time-domain coherent phonon signal at several representative temperatures. Figure \ref{temperature_dependence}(b) shows the two-dimensional color map of the normalized FT spectrum for each temperature. As shown in Fig. \ref{temperature_dependence}(b), no emergence of new phonon modes or the disappearance of existing ones was confirmed, consistent with the absence of other structural phases in the measured low temperature region and the pump fluence was sufficiently smaller than the fluence required to induce the phase transition ($\sim$ several mJ/cm$^{2}$) \cite{Akei2025switching}. To discuss the temperature dependence of coherent phonons, the time-domain signals at each temperature were fitted with the sum of damped harmonic oscillations:
\begin{align}
f(t)=\sum_{i} A_ie^{-\Gamma_i t}\cos(2\pi\omega_i t+\phi_i),
\label{damped}
\end{align}
where $A_i, \Gamma_i,\omega_i$ and $\phi_i$ are the amplitude, the decay rate, the frequency, and the initial phase of each mode, respectively. 
We have chosen three representative intralayer phonon modes $\omega_2,\,\omega_6$ and $\omega_7$, since they have large amplitudes and non-degenerated. The interlayer shear mode, $\omega_1$, showed the largest experimental errors 
as well as exhibited conventional anharmonic phonon decay \cite{he2016coherent} and therefore we do not examine $\omega_1$ in the present study. 
As shown in Fig. \ref{temperature_dependence}(a), the fitting results reproduce the data extremely well.

\begin{figure}
\makeatletter
\renewcommand{\thefigure}{%
\thesection \arabic{figure}}
\@addtoreset{figure}{section}
\makeatother
	\includegraphics[width=8.6cm]{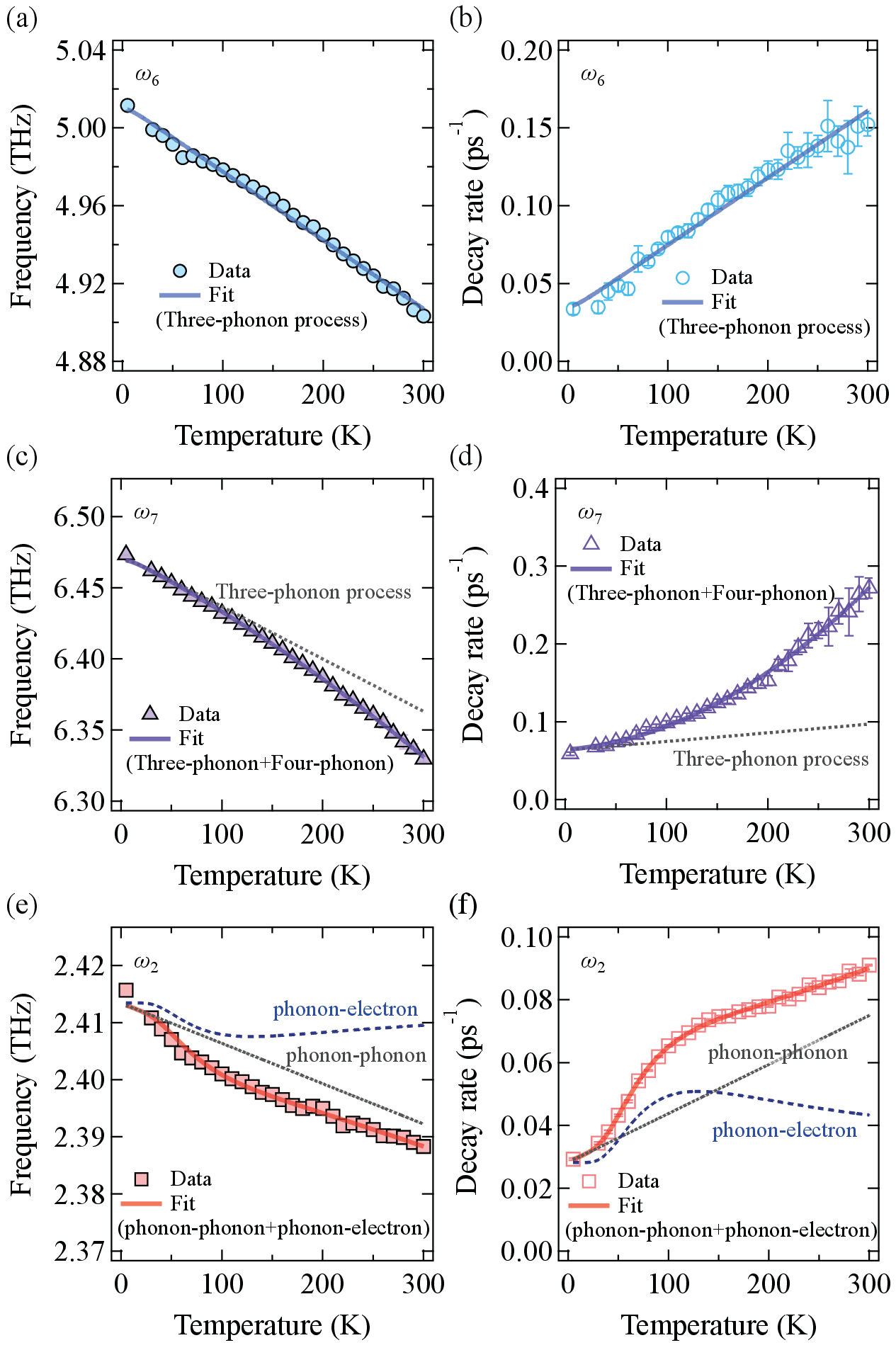}
	\caption{(a), (b) Temperature dependence of the frequency and decay rate for the $\omega_6$ phonon mode. The blue solid lines represent the fit using anharmonic phonon-phonon scattering model,
    Eqs. (2) and (3). 
    (c), (d) Temperature dependence of the frequency and decay rate for the $\omega_7$ phonon mode. The purple solid lines represent the fit using anharmonic phonon-phonon scattering model, Eqs. (4) and (5). 
    (e), (f) Temperature dependence of the frequency and decay rate for the $\omega_2$ phonon mode. 
    The red solid lines represent the fit using Eqs. (6) and (7).
    The gray dotted lines represent the anharmonic phonon-phonon scattering process, while the blue dotted lines represent the phonon-electron scattering process. The error bars represent the standard deviations obtained from the fitting procedure.}
	\label{decay_freq}
\end{figure}

We evaluated the temperature dependence of the frequency and decay rate for each phonon mode using parameters obtained by fitting using Eq. (\ref{damped}). As discussed below, the differences in the scattering mechanisms for each mode are attributed to their energy scales.
Figure \ref{decay_freq}(a) and \ref{decay_freq}(b) show the temperature dependence of the oscillation frequency and the decay rate of the $\omega_6$ phonon mode. As the temperature increases, the frequency decreases monotonically, and the decay rate increases monotonically. This behavior is predicted by anharmonic phonon-phonon scattering due to the anharmonic terms of the lattice potential \cite{klemens1966anharmonic}. The most general mechanism for the decay of optical phonons is the three-phonon process \cite{hase2015anharmonic}, in which an optical phonon decays into two underlying acoustic phonons of the same frequency with opposite wave vectors. The temperature dependence of the frequency for the three-phonon process is described by the following Klemens model \cite{klemens1966anharmonic}:
\begin{align}
\omega(T)=\omega_0+\omega_{3\mathrm{ph}}\Bigg\{1+\frac{2}{\exp\big(\frac{\hbar\omega_0}{2k_BT}\big)-1}\Bigg\},
\label{three-phonon_w}
\end{align}
where $\omega_0$ is the frequency at 0 K, $\omega_{3\mathrm{ph}}$ is the fitting parameter, and $k_B$ is the Boltzmann constant. Since phonons are of bosonic nature, they are governed by the Bose-Einstein distribution function. As shown in Fig. \ref{decay_freq}(a), the redshift in the frequency of the $\omega_6$ mode due to the increase in temperature can be well reproduced by the three-phonon process. Similarly, the temperature dependence of the decay rate is given by the following equation \cite{klemens1966anharmonic}:
\begin{align}
\Gamma(T)=\Gamma_0+\Gamma_{3\mathrm{ph}}\Bigg\{1+\frac{2}{\exp\big(\frac{\hbar\omega_0}{2k_BT}\big)-1}\Bigg\},
\label{three_phonon_g}
\end{align}
where $\Gamma_0$ is the decay rate at 0 K and $\Gamma_{3\mathrm{ph}}$ is the fitting parameter. The well-fitting results obtained from Eqs. (\ref{three-phonon_w}) and (\ref{three_phonon_g}) indicate that the scattering of the $\omega_6$ mode is dominated by the three-phonon process.

Next, we focus on the $\omega_7$ mode. As shown in Fig. \ref{decay_freq}(c) and \ref{decay_freq}(d), the observed temperature dependence can become nonlinear and 
cannot be fully explained by the three-phonon process. Therefore, we use the Balkanski model, which is an extended Klemens model, and includes a higher-order four-phonon process (i.e., an optical phonon decays into three acoustic phonons) in addition to the three-phonon process. The temperature dependence of the phonon frequency and decay rate is expressed by the following equations \cite{balkanski1983anharmonic}:
\begin{align}
&\omega(T)=\omega_0+\omega_{3\mathrm{ph}}\Bigg\{1+\frac{2}{\exp\big(\frac{\hbar\omega_0}{2k_BT}\big)-1}\Bigg\}\notag
\\&+\omega_{4\mathrm{ph}}\Bigg[1+\frac{3}{\exp\big(\frac{\hbar\omega_0}{3k_BT}\big)-1}+\frac{3}{\big\{\exp\big(\frac{\hbar\omega_0}{3k_BT}\big)-1\big\}^2}\Bigg],
\label{four_phonon_w}
\end{align}

\begin{align}
&\Gamma(T)=\Gamma_0+\Gamma_{3\mathrm{ph}}\Bigg\{1+\frac{2}{\exp\big(\frac{\hbar\omega_0}{2k_BT}\big)-1}\Bigg\}\notag\\&+\Gamma_{4\mathrm{ph}}\Bigg[1+\frac{3}{\exp\big(\frac{\hbar\omega_0}{3k_BT}\big)-1}+\frac{3}{\big\{\exp\big(\frac{\hbar\omega_0}{3k_BT}\big)-1\big\}^2}\Bigg],
\label{four_phonon_g}
\end{align}
where $\omega_0$ and $\Gamma_0$ are the frequency and the decay rate at 0 K, $\omega_{3\mathrm{ph}}$, $\omega_{4\mathrm{ph}}$, $\Gamma_{3\mathrm{ph}}$ and $\Gamma_{4\mathrm{ph}}$ are the fitting parameters. 

The fitted results by Eqs. (\ref{four_phonon_w}) and (\ref{four_phonon_g}) well reproduce the experimental results shown in Figs. \ref{decay_freq}(c) and \ref{decay_freq}(d), with the global fit parameters (in THz) of 
$\omega_{3\mathrm{ph}}=-0.00457\pm0.000403$, $\omega_{4\mathrm{ph}}=-0.0000320\pm0.00000894$, $\Gamma_{3\mathrm{ph}}=0.00138\pm0.000401$ and $\Gamma_{4\mathrm{ph}}=0.000176\pm0.00000896$.
The fraction of each process obtained indicate that a contribution of the 
four-phonon process plays a minor role of $\sim$10\%. 
One possible reason for the inclusion of the four-phonon process is that $\omega_7$ ($\sim$ 6.3 THz) has relatively high energy. 
Based on phonon dispersion, it would be difficult to distribute its energy only through a three-phonon process to satisfy energy and momentum conservations \cite{jiang2016raman}.
Consequently, it can be concluded that the four-phonon process as well as the three-phonon process contribute to the scattering mechanism of the $\omega_7$ mode.

Finally, the temperature dependence of the frequency and decay rate for $\omega_2$ mode is shown in Figs. \ref{decay_freq}(e) and \ref{decay_freq}(f), respectively. Unlike the $\omega_6$ and $\omega_7$ modes, $\omega_2$ mode exhibited non-monotonic behavior with a critical point ($\sim$100 K), which cannot be explained by the anharmonic decay model. Such a critical point is sometimes observed in phase transition materials \cite{cao2025enhanced}, but this is not the case here because the phase transition in WTe$_2$ occurs at 565 K \cite{tao2020t}. The anomalous behavior indicates the existence of other scattering mechanisms. Thus, in addition to phonon-phonon scattering, we consider the contribution of the phonon-electron scattering process, in which an optical phonon decays into an electron-hole pair through interband transition \cite{yu2023anomalous,osterhoudt2021evidence,coulter2019uncovering,yang2021evidence}. 

\begin{figure}
\makeatletter
\renewcommand{\thefigure}{%
\thesection \arabic{figure}}
\@addtoreset{figure}{section}
\makeatother
	\includegraphics[width=6cm]{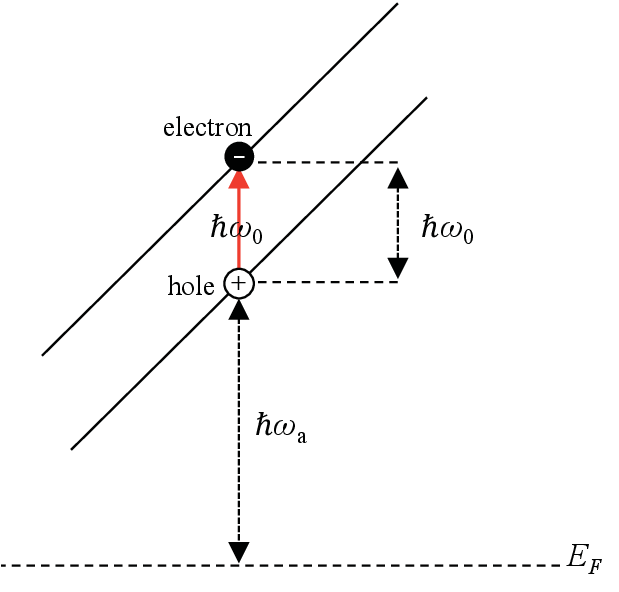}
	\caption{Schematic illustration of the interband transition (red arrow) via optical phonon with energy of $\hbar\omega_{\mathrm {0}}$ for parallel electronic bands.
    $\hbar\omega_{a}$ represents the energy difference between the initial electronic state and the Fermi energy $E_F$.}
	\label{band}
\end{figure}

When full considerations for the electronic bands are introduced, according to the Fermi golden rule, the temperature dependence of the decay rate (or imaginary part of the phonon self-energy) via electron-phonon scattering can be given by \cite{bonini2007phonon}
\begin{align}
\Gamma_{\mathrm{ph-e}}(T)&=\frac{4\pi}{N_k}\sum_{\textbf{k}, i, j}|g_{\textbf{(k+q)}j,\,\textbf{k}i}|^2[f_{\textbf{k}i}(T)-f_{\textbf{(k+q)}j}(T)]\notag\\&\times\delta[\epsilon_{\textbf{k}i}-\epsilon_{\textbf{(k+q)}j}+\hbar\omega_{\textbf{q}}],
\label{Im_selfenergy}
\end{align}
where, a phonon with wave vector \textbf{q} excites an electronic state $\ket{\textbf{k}, i}$ into the state $\ket{\textbf{k+q}, j}$. The sum is on $N_k\textbf{k}$ vectors, $\omega_{\textbf{q}}$ is the phonon frequency, $f_{\textbf{k}i}(T)$ is the Fermi-Dirac occupation, and $\delta$ is the Dirac delta function. $g_{\textbf{(k+q)}j, \textbf{k}i}$ is the electron-phonon coupling matrix element given by, \cite{piscanec2004kohn}
\begin{align}
g_{\textbf{(k+q)}j,\textbf{k}i}=\bra{\textbf{k+q}, j}\Delta V_{\textbf{q}}[\Delta n_{\textbf{q}}]\ket{\textbf{k}, i}\sqrt{\hbar/(2M\omega_{\textbf{q}})},
\label{EPC}
\end{align}
where $\Delta V_{\textbf{q}}$ and $\Delta n_{\textbf{q}}$ are the derivative of the Kohn-Sham potential \cite{Kohn-Sham1965} and of the charge density with respect to a displacement along the normal coordinate of the phonon, and $M$ is the atomic mass. The simple forms can be obtained from Eq. (\ref{Im_selfenergy}) if we assume the parallel bands (Fig. \ref{band}), so that $\hbar\omega_{\textbf{q}}=\epsilon_{\textbf{(k+q)}j}-\epsilon_{\textbf{k}i}$ can be satisfied for all \textbf{k}.

Thus, the temperature dependence of frequency and decay rate, comprising the phonon-phonon scattering and phonon-electron scattering terms, can be given by the following phenomenological model \cite{hou2025probing}:
\begin{align}
&\omega(T)=\omega_0+\omega_{3\mathrm{ph}}\Bigg\{1+\frac{2}{\exp\big(\frac{\hbar\omega_0}{2k_BT}\big)-1}\Bigg\}\notag\\&+\omega_{\mathrm{ph-e}}\Bigg[\frac{1}{\exp\big(\frac{\hbar\omega_a}{k_BT}\big)+1}-\frac{1}{\exp\big\{\frac{\hbar(\omega_a+\omega_0)}{k_BT}\big\}+1}\Bigg],
\label{ph_e_w}
\end{align}

\begin{align}
&\Gamma(T)=\Gamma_0+\Gamma_{3\mathrm{ph}}\Bigg\{1+\frac{2}{\exp\big(\frac{\hbar\omega_0}{2k_BT}\big)-1}\Bigg\}\notag\\&+\Gamma_{\mathrm{ph-e}}\Bigg[\frac{1}{\exp\big(\frac{\hbar\omega_a}{k_BT}\big)+1}-\frac{1}{\exp\big\{\frac{\hbar(\omega_a+\omega_0)}{k_BT}\big\}+1}\Bigg],
\label{ph_e_g}
\end{align}
where $\omega_0$ and $\Gamma_0$ are the frequency and the decay rate at 0 K, $\omega_{\mathrm{ph-ph}}$, $\omega_{\mathrm{ph-e}}$, $\Gamma_{\mathrm{ph-ph}}$ and $\Gamma_{\mathrm{ph-e}}$ are the fitting parameters, $\hbar\omega_a$ is the energy of the initial electron state relative to the Fermi energy. The third term on the right-hand side describes phonon-electron scattering, namely an optical phonon decay into an electron-hole pair via phonon-mediated electron excitation as shown in Fig. \ref{band}. For this process to occur, the initial state must be occupied, and the final state must be empty. Thus, the scattering probability is expressed as the difference between the Fermi distribution function of the initial state and that of the final state \cite{osterhoudt2021evidence}. 
As shown in Fig. \ref{decay_freq}(e) and (f), the global fitting results for temperature dependence using the model considering both processes reproduced the data very well and support this interpretation. 

The reason why phonon-electron scattering contributes only to the $\omega_2$ mode is that the electron-phonon coupling of the $\omega_2$ mode is stronger than that of other modes. In fact, the electron-phonon coupling constant $g$ corresponds to the degree of band shift caused by lattice displacement, and according to tr-ARPES study \cite{hein2020mode}, a significant band shift observed at 2.4 THz, suggesting that the electron-phonon coupling of the $\omega_2$ mode is strong. In addition, from Eq. (\ref{EPC}), we know $g\propto\sqrt{1/\omega}$ if we assume the parallel bands. Therefore, excluding the shear mode ($\omega_1$)
which is governed by anharmonic phonon decay \cite{he2016coherent}, electron-phonon coupling matrix element $g$ is largest for the lowest phonon mode, 2.4 THz in the present case.
Furthermore, the phonon lifetime of the $\omega_2$ mode is extremely long ($\sim$ 34 ps at 4.6 K and $\sim$ 11 ps at 300 K) compared to that of other modes, indicating that the effect of anharmonic phonon-phonon scattering can be smaller than that of phonon-electron scattering below the Debye temperature ($\approx$133 K) \cite{CALLANAN1992627}. 
Based on the above careful considerations, it can be concluded that the contribution from phonon-electron scattering plays an important role in the temperature dependence of the decay rate and frequency for the $\omega_2$ mode.

From another point of view, it is also considered that, due to changes in the electronic structure with varying temperature (such as the Lifshitz transition), there exist phonon-electron scattering pathways in addition to phonon-phonon scattering. The Lifshitz transition is characterized by abrupt topological changes in the Fermi surface. It has been observed in materials such as ZrTe$_{5}$ \cite{zhang2017electronic}, HfTe$_{5}$ \cite{zhang2017temperature}, NiTe$_{2}$ \cite{cheng2022ultrafast} and MoTe$_{2}$ \cite{beaulieu2021ultrafast, lu2025probing}. In particular, it was demonstrated in NiTe$_2$ that the Lifshitz transition leads to an abrupt change in the temperature dependence of the phonon decay rate and frequency \cite{cheng2022ultrafast}. In WTe$_{2}$, hole pockets emerge below the transition temperature ($\sim$160 K \cite{wu2015temperature, sruthi2022interband}) due to a shift in chemical potential \cite{luo2025nonlinear}. This may influence the probability of phonon scattering through the formation of electron-hole pairs. Although the proposed transition temperature (160 K) deviates from the critical temperature we observed ($\sim$100 K in Fig. \ref{decay_freq}), it has been revealed that the transition temperature varies with the thickness of the sample. Specifically, thicker samples (> 10 layers) exhibit a Lifshitz transition at lower temperature ($\sim$100 K \cite{luo2025nonlinear}), which coincides with our observation. 
Therefore, the temperature dependence of the decay rate for the $\omega_2$ mode which exhibits an obvious change in slope around 100 K [Fig. \ref{decay_freq}(f)], may be induced by the Lifshitz transition.

\begin{figure}
\makeatletter
\renewcommand{\thefigure}{%
\thesection \arabic{figure}}
\@addtoreset{figure}{section}
\makeatother
	\includegraphics[width=8.6cm]{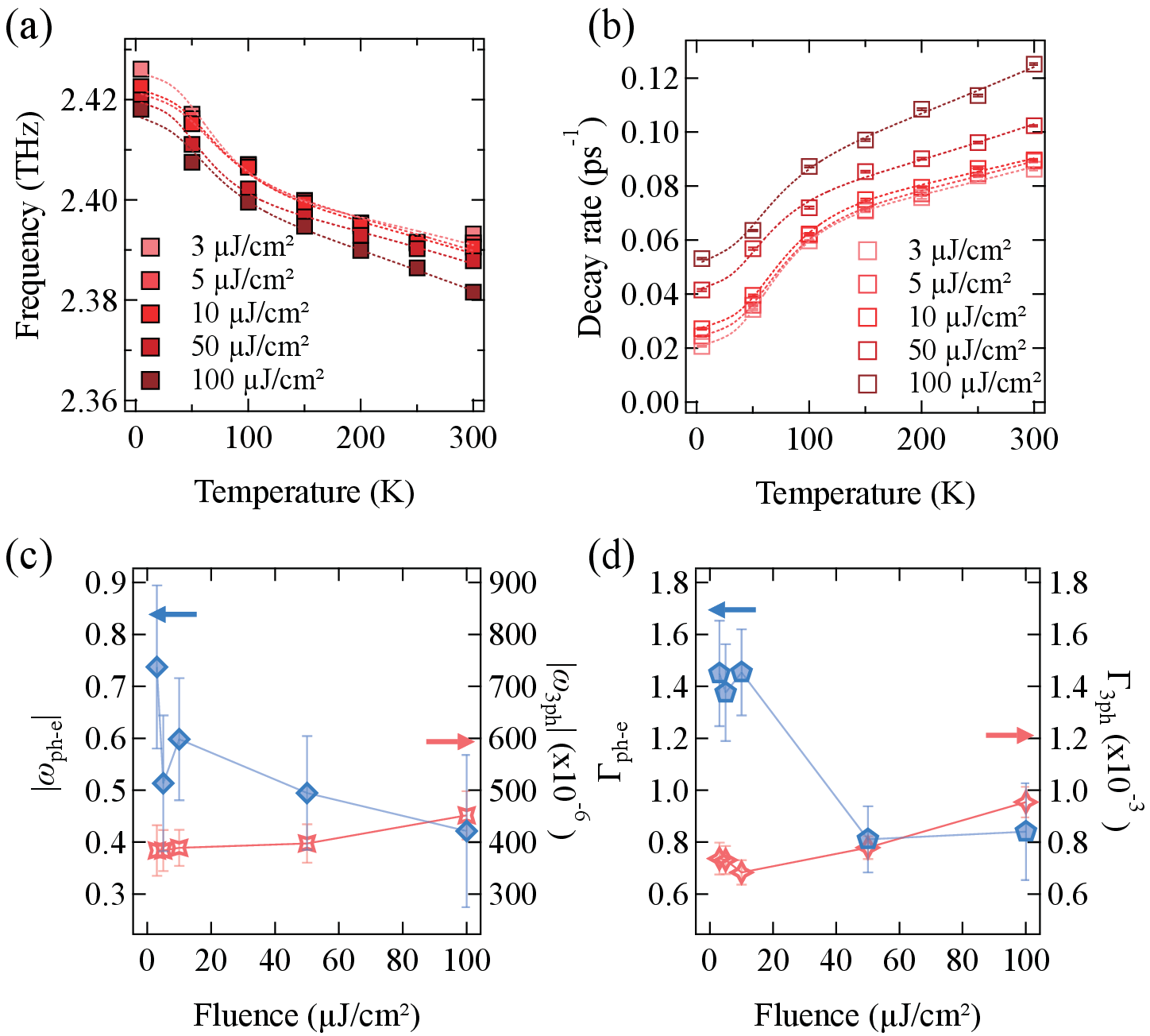}
	\caption{(a), (b) Temperature dependence of the frequency and decay rate for the $\omega_2$ phonon mode in the range of pump fluence from 3 to 100 $\mu$J/cm$^{2}$. The dotted lines show the fits obtained from Eqs. (\ref{ph_e_w}) and (\ref{ph_e_g}). (c), (d) Fluence dependence of fit parameters $\omega_{3\mathrm{ph}}$, $\omega_{\mathrm{ph-e}}$, $\Gamma_{3\mathrm{ph}}$, and $\Gamma_{\mathrm{ph-e}}$. The error bars represent the standard deviations obtained from the fitting procedure.}
	\label{fluence_dependence}
\end{figure}

To further understand the role of phonon-electron scattering in the $\omega_2$ phonon mode, we varied the pump fluence from 3  to 100 $\mu$J/cm$^{2}$ and measured the temperature dependence of coherent phonons. Figures \ref{fluence_dependence}(a) and \ref{fluence_dependence}(b) show the temperature dependence of the frequency and decay rate of the $\omega_2$ mode for each fluence, together with the results of the fit using Eqs. (\ref{ph_e_w}) and (\ref{ph_e_g}) represented by the dotted line. As mentioned above, the fitting results for each fluence reproduced the experimental data well. 
The slight deviations between the frequency and decay rate of the $\omega_2$ mode at 10 $\mu$J/cm$^{2}$ shown in Figs. \ref{decay_freq}(e) and \ref{decay_freq}(f) and the corresponding data presented in Figs. \ref{fluence_dependence}(a) and \ref{fluence_dependence}(b) are due to different measurement runs, however, they do not affect the conclusions in the present study.

The fluence dependence of the magnitudes of the coefficients for the phonon-phonon scattering term ($\omega_{3\mathrm{ph}}$ and $\Gamma_{3\mathrm{ph}}$) and the phonon-electron scattering term ($\omega_{\mathrm{ph-e}}$ and $\Gamma_{\mathrm{ph-e}}$) in Eqs. (\ref{ph_e_w}) and (\ref{ph_e_g}) is shown in Figs. \ref{fluence_dependence}(c) and \ref{fluence_dependence}(d). These represent the fluence dependence of the contributions of each scattering process. Interestingly, the fluence dependence of the coefficients for the phonon-phonon scattering term and the phonon-electron scattering term exhibits opposite trends. 
The former indicates a small increase as the fluence increases from 3 to 100 $\mu$J/cm$^{2}$, whereas the latter largely decreases. The contribution of phonon-phonon scattering slightly increases with increasing excitation fluence possibly due to modification of the lattice potential under higher excitation \cite{Hase2002Anharmonic}.
In contrast, the decrease in the phonon-electron scattering as the fluence increases can be interpreted as screening of the electron-phonon coupling by photogenerated carriers \cite{Sohier2016Screening}. As fluence increases, the carrier density increases, and the interaction between electrons and phonons is weakened. Although Ref. \cite{Sohier2016Screening} mainly discuss the screening effect in monolayer TMDs, they also reported that Fr\"ohlich interaction was screened due to the $\propto 1/|\textbf{q}|$ factor even for the case of bulk TMDs. In addition, according to a different study \cite{pan2025momentum}, it is claimed that the electron-phonon coupling in bulk TMDs is suppressed by carrier screening. Therefore, the mechanism of screening of the electron-phonon coupling by photogenerated carriers is likely to apply to bulk $T_d$-WTe$_2$ as well.

We point out that WTe$_2$ system may be suitable for observing and controlling dynamical Lifshitz transition \cite{beaulieu2021ultrafast}, since the contribution from the phonon-phonon scattering and phonon-electron scattering terms can be tuned by the photogenerated carrier density as demonstrated in Fig. \ref{fluence_dependence}. In addition, phonon-electron scattering plays a vital role in phonon scattering mechanisms in other 2D Weyl semimetals \cite{yu2023anomalous} and even 3D metallic systems \cite{yang2021evidence}, and therefore our approach will be applicable to a wide range of quantum materials.

In conclusion, we investigated the phonon scattering dynamics of WTe$_2$ by femtosecond transient optical pump-probe spectroscopy in the wide temperature range from 4.6 to 300 K. 
We evaluated the temperature dependence of the phonon frequency and decay rate for the selected three phonon modes, whose differences in the scattering mechanisms are attributed to their energy scales.
In two high-frequency modes, both frequency and decay rate as a function of temperature can be well fitted with the anharmonic phonon-phonon scattering model. In contrast, the low-frequency mode with the frequency of 2.4 THz can be explained using the phonon-electron scattering model in addition to the anharmonic phonon-phonon scattering model. 
Furthermore, the measurements of fluence dependence revealed that the contribution from phonon-electron scattering decreases with increasing excitation fluence, which is interpreted to arise from the screening effect caused by the increase in carrier density. 
Our results provide a framework for understanding the mechanisms for phonon scattering in Weyl semimetals through coherent phonon spectroscopy, and the findings of the phonon-electron scattering path 
(possibly the Lifshitz transition)
pave the way for further exploration of the electronic structure and transport properties in a wide range of quantum materials.

\nocite{*}

\section*{Acknowledgement}
This work was supported by KAKENHI from the Japan Society for the Promotion of Science (JSPS) (Grant Number. 25H00849).

\appendix

\section{Heating effect}
The lattice temperature increase $\Delta T$ due to laser irradiation at the initial temperature $T_0$ can be roughly estimated by the following equation \cite{harter2018evidence}:
\begin{align}
\frac{S\lambda\rho}{M}\int^{T_0+\Delta T}_{T_0}C(T)dT=(1-R)FS
\label{heating effect}
\end{align}
where $S,\,\lambda,\,\rho,\,M,\,C(T),\,R$ and $F$ are the excitation area, the optical penetration depth ($\approx$ 30 nm at 830 nm from our own measurement), the density (9.43 g/cm$^{3}$), the molar mass (439.05 g/mol), the temperature dependent lattice heat capacity, the optical reflectivity (about 0.45 at 830 nm from our own measurement) and the incident pump fluence. For the pump fluence of 10 $\mu$J/cm$^{2}$, the value of $\Delta T$ was estimated to be $\sim$ 21 K at $T_0=6$ K, and < 4 K at $T_0\geqq 40$ K. For the pump fluence of 100 $\mu$J/cm$^{2}$, the value of $\Delta T$ was estimated to be $\sim$ 52 K and $\sim$ 12 K, respectively, at $T_0=6$ K and $T_0=300$ K. At the pump fluence of 100 $\mu$J/cm$^{2}$, the laser irradiation exhibits non-negligible heating effect, but this is not significant at the pump fluence of 10 $\mu$J/cm$^{2}$. As the thermal diffusion has not been taken into account here, the realistic temperature increase is considered to be even lower.

\bibliography{apssamp}
\end{document}